\documentclass[twocolumn,10pt]{revtex4}%
\usepackage[paperwidth=210mm,paperheight=297mm,centering,hmargin=2cm,vmargin=2.5cm]{geometry}
\usepackage{amsfonts}
\usepackage{amsmath}
\usepackage{amssymb}
\usepackage{bm}
\usepackage{graphicx}
\usepackage{color}
\usepackage{soul}
\usepackage{epsfig}%
\usepackage[dvipsnames]{xcolor} 
  \definecolor{bleu_cite}{RGB}{0,0,255}
\usepackage[colorlinks=true,allcolors = black,citecolor=bleu_cite,  ]{hyperref} 
\usepackage{natbib}
\usepackage{siunitx}

\def\fd{\textcolor{black}}
\def\fdd{\textcolor{black}}

\usepackage[normalem]{ulem} 

\graphicspath{{figures/}}

\begin{document}
\title{\fd{Extended Bose-Hubbard model with dipolar excitons}}

\author{C. Lagoin$^1$, U. Bhattacharya$^2$,  T. Grass$^2$, R. W. Chhajlany$^{3}$, T. Salamon$^2$, K. Baldwin$^4$, L. Pfeiffer$^4$, M. Lewenstein$^{2,5}$, M. Holzmann$^6$ and F. Dubin$^{1,\dag,\star}$} 
\affiliation{$^1$ Institut des Nanosciences de Paris, CNRS and Sorbonne Universit{\'e}, Paris, France}
\affiliation{$^2$ ICFO – Institut de Ciencies Fotoniques, The Barcelona Institute of Science and Technology, Castelldefels, Spain}
\affiliation{$^3$ Institute of Spintronics and Quantum Information, Faculty of Physics, Adam Mickiewicz University, 61-614 Pozna{\'n}, Poland}
\affiliation{$^4$ PRISM, Princeton Institute for the Science and Technology of Materials, Princeton University, Princeton, NJ 08540, USA}
\affiliation{$^5$ ICREA, Pg. Lluís Companys, Barcelona, Spain}
\affiliation{$^6$ Univ. Grenoble Alpes, CNRS, LPMMC, 38000 Grenoble, France}
\affiliation{$\star$: Present address CRHEA -- CNRS, Valbonne, France}
\affiliation{$\dag$: francois$\_$dubin@icloud.com (corresponding author)}

\begin{abstract}
\textbf{The Hubbard model constitutes one of the most celebrated theoretical frameworks of condensed-matter physics. It describes strongly correlated phases of interacting quantum particles confined in lattice potentials \cite{Salomon_2010,Arovas_2022}. For bosons, the Hubbard Hamiltonian has been deeply scrutinised for short-range on-site interactions \cite{Greiner_2002,Chin_2009,Greiner_Mott,Bloch_2010}. \fd{However, accessing longer-range couplings has remained elusive experimentally \cite{Ferlaino_2016}. This marks the frontier towards the extended Bose-Hubbard Hamiltonian that allows insulating ordered phases at fractional lattice fillings \cite{Batrouni_95,Hebert_01,Maciej_2002,Baranov_2011,Dutta_2015}. Here we implement this Hamiltonian by confining semiconductor dipolar excitons in an artificial two-dimensional square lattice. Strong dipolar repulsions between nearest neighbouring lattice sites then stabilise an insulating state at half filling. This characteristic feature of the extended Bose-Hubbard model exhibits signatures theoretically expected  for a checkerboard spatial order. Our work thus highlights that dipolar excitons enable controlled implementations of boson-like arrays with strong off-site interactions, in lattices with programmable geometries and over 100 sites.}}
\end{abstract}

\maketitle

\fd{\textbf{Introduction:}} The extended Bose-Hubbard (BH) Hamiltonian is controlled by three main physical parameters -- the strength of on-site interactions $U$, the  tunnelling strength $t$, and the interaction strength between nearest neighbouring lattice sites $V$ \cite{Salomon_2010}. While the standard regime where $V$ plays a negligible role has been thoroughly studied \cite{Greiner_2002,Chin_2009,Greiner_Mott,Bloch_2010}, exploring the situation where $V$ controls the many-body ground-state continues to seriously challenge experimental research in condensed-matter physics \cite{Ferlaino_2016}. In this regime, theoretical works have predicted that long-range interactions stabilise quantum phases spontaneously breaking the lattice symmetry \cite{Batrouni_95,Hebert_01,Baranov_2011,Dutta_2015,Maciej_2002,Trefzger_2010,Capogrosso_2010}, like density waves manifesting checkerboard (CB) or stripe solids. Such phases have been observed for fermionic systems \cite{Tranquada_2015,Wise_2008,Jin_2021} while their implementation for bosons still constitutes a long-lasting goal.

Theoretically, it is established that dipolar bosons in a lattice provide an ideal platform to implement the extended BH model \cite{Maciej_2002,Baranov_2011,Dutta_2015}. Here, we experimentally follow this path by confining semiconductor dipolar excitons in an artificial two-dimensional square lattice. We then evidence two insulating phases, at unity and half-filling of the lattice sites. The former case marks the Mott insulator (MI) regime where lattice sites are all occupied by a single exciton \cite{Lagoin_2022}. On the other hand, an incompressible region at half-filling points towards an exciton density wave, which spontaneously breaks the lattice symmetry, favoured by nearest-neighbour (NN) repulsions. For our physical parameters mean-field and exact diagonalisation calculations predict that this phase corresponds to a CB solid. Importantly, \fd{we directly support this expectation, by measuring the thermal melting of both insulating phases that quantitatively agree with theoretical predictions. }

\fd{\textbf{Dipolar excitons in a lattice:}} In recent years, different techniques for engineering tunable lattice potentials in excitonic systems have been developed, including optical \cite{Lackner_2021} or Moir{\'e} lattices \cite{moire_1,moire_2,moire_3,moire_4,Kennes_2021,Shan_2022}. As illustrated in Fig.1.a, here we follow the approach devised in Refs. \cite{Butov_lattice,Lagoin_2020,Lagoin_2021,Lagoin_2022} and polarise an array of gate electrodes deposited at the surface of a field-effect device embedding a GaAs double quantum well. These electrodes imprint a sinusoidally varying electric field, perpendicular to the plane of the two quantum wells where electrons (black balls) and holes (white balls) are spatially separated to realise dipolar excitons (red arrows) \cite{Combescot_ROPP,High_2009,Winbow_2011,Shilo_2013}. Excitonic dipoles being all aligned with the applied electric field, dipolar excitons are confined in an artificial lattice with 250 nm period where they interact through a quasi-long-range repulsive dipolar potential. This provides a unique environment to emulate the extended Bose-Hubbard Hamiltonian.

In our experiments excitons are optically injected, using a \SI{100}{\nano\second} long laser excitation repeated at \SI{1}{\mega\Hz}. The laser has a rectangular profile at the surface of our device, with around (10x5) $\mu$m$^2$ dimension, its average power $P$ controlling the mean exciton density in the lattice potential. We analyse stroboscopically the photoluminescence (PL) emitted in a (3x2.5) $\mu$m$^2$ area at the center of the illuminated region, \SI{300}{\nano\second} after termination of the loading laser pulse and at bath temperatures $T$ as low as 330 mK. Moreover, let us underline that a single measurement refers to an acquisition lasting typically 30 to 60 seconds, thereby accumulating 3 to 6$\cdot$10$^7$ experimental realisations (Methods). In the following, we discuss experimental results obtained by statistically analysing 10 measurements for every experimental conditions.


The states accessible to dipolar excitons in the lattice are directly visualised in the PL spectrum emitted in the very dilute limit, i.e. when the mean density $\bar{n}$ is around 0.2 exciton per site. Indeed, Fig.1.d shows that the spectrum then consists of two peaks separated by \SI{150}{\micro\eV}. This splitting matches the energy separation theoretically calculated between the two Wannier states (WS) confined in the lattice (Fig.1.b), the lattice depth being around 250 $\mu$eV (Methods). Thus, we quantitatively reproduce the PL spectrum by adding two lorentzian-like lines separated by \SI{150}{\micro\eV}, each line profile being given by the spectral response of our imaging spectrometer (with 150 $\mu$eV full-width-at-half-maximum). In Fig.1d, we then only adjust the two lines amplitudes that reflect the fraction of excitons occupying the corresponding WS (orange and green). On the other hand, in a denser regime, when $\bar{n}\sim$1.3, Fig.1.e highlights that a second emission (blue area) emerges. It  occurs at too high energies to be possibly attributed to lattice confined states. In fact, it signals excitons occupying the continuum states accessible above the lattice potential (blue in Fig.1.b). The energy splitting between this contribution and the lattice one, around 500 $\mu$eV, provides an order of magnitude of the dipolar interaction strength between excitons.

\fd{\textbf{Extended BH model:}} Following the approach detailed in Ref.~\cite{Lagoin_2022,Gotting_2022}, we computed the parameters of the extended Bose-Hubbard Hamiltonian. We assumed that the two WS (1 and 2 in Fig.1.b) correspond to $s$ and $p$ orbitals respectively \cite{Hofstetter_2012}. Then, we deduced that on-site interactions have a strength greatly exceeding the lattice depth, with for instance $U_{1,1}\sim$ \SI{1}{\milli\eV} for excitons populating WS 1 (Methods B). As a result, lattice sites cannot be doubly occupied, since on-site interactions easily ``expel" excitons into the continuum. Moreover, the interaction strength $V$ between excitons in nearest neighbouring lattice sites largely exceeds the tunnelling strength $t$ for both WS (Methods), as necessary to stabilise symmetry-breaking phases \cite{Dutta_2015}.

Figure 1.c presents the phase diagram calculated in the ($\mu,T$) parameter space, $\mu$ denoting the chemical potential. The diagram was computed for an extended two-band BH Hamiltonian in the mean-field approximation, combining measured and calculated values for $U$, $V$ and $t$ for each WS (Methods B-D). The accuracy of this approach was confirmed by exact diagonalization (ED) of the full Hamiltonian. Figure 1.c first shows that a MI is energetically favoured at unity filling of the lattice sites (purple lobe). As reported recently \cite{Lagoin_2022}, this phase is marked by a minimised compressibility and by excitons uniformly occupying the same WS in every lattice site. Moreover, ED (\fd{extended data Fig.1}) and mean-field theory strikingly predict that, at half-filling of the lattice sites, $V$ is sufficiently large to stabilize a checkerboard (CB) solid (orange lobe). For this incompressible density wave the exciton distribution is such that NN interactions are fully avoided. Finally, in Fig.1.c we recover that MI and CB are surrounded by a normal fluid phase (NF -- gray region). Indeed, superfluid properties are only accessible at temperatures below 20 -- 30 mK for $\bar{n}\sim1$.

\fd{\textbf{Incompressible phases in the lattice:}} To detect the buildup of insulating phases in the lattice, we measured the exciton compressibility $\kappa$. For that we 
monitored statistically the maximum of the PL spectrum ($A_\mathrm{max}$). Precisely, we computed the average $\overline{A_\mathrm{max}}$ and standard deviation $\sigma(A_\mathrm{max})$, which directly quantify $\kappa$, since $\sigma(A_\mathrm{max})/\overline{A_\mathrm{max}}$  is proportional to $(\kappa k_BT)^{1/2}$   according to the fluctuation-dissipation theorem \cite{Chin_2009}.

Figure 2.a presents the variation of $\sigma(A_\mathrm{max})/\overline{A_\mathrm{max}}$ at $T=$ \SI{330}{\milli\kelvin}, as a function of the average power of the loading laser $P$. For two specific excitations, $P=$ 8 and \SI{17}{\nano\watt} (center of orange and purple regions), Fig.2.a shows that $\kappa$ is strongly decreased compared to the level of Poissonian fluctuations (gray region).  Furthermore, for $P=$ \SI{17}{\nano\watt} we expect that $\bar{n}\sim1$ (Methods). Noting that PL intensities differ by around two-fold between $P=8$ and $P=$ \SI{17}{\nano\watt} (Fig.2.b-d), we deduce that $\bar{n}=1/2$ for the former excitation. Accordingly, Fig.2.a signals two insulating phases, which extend over 100 lattice sites according to spatially resolved PL intensity and intensity fluctuations (\fd{extended data Fig.2}). Importantly, we verified that the emergence of the incompressible states at $\bar{n}=$ 1 and 1/2 does not depend on the region of the lattice explored experimentally (\fd{extended data Fig.3}).

Figure 2.b and 2.d compare the PL spectra radiated at $\bar{n}$=1 and 1/2. We first note that for the latter case (Fig.2.b) the maximum of the PL lies at the energy of the 1$^{st}$ WS (see left vertical line), while for $\bar{n}$=1 it does not coincide with any WS energy (Fig.2.d). Instead, the PL maximum lies at $\Delta\sim$100 $\mu$eV above the energy of the 1$^{st}$ WS. Since we theoretically expect that for $\bar{n}$=1/2 and 1 excitons occupy the 1$^{st}$ WS and realize CB and MI phases respectively, we deduce that $\Delta=4V_{1,1}$, $V_{1,1}$ denoting the strength of NN interactions for the 1$^{st}$ WS. Figure 2 thus yields $V_{1,1}=25\pm5$ $\mu$eV.

We have scrutinised the buildup of the two insulating phases by modelling the PL spectra emitted for every average filling $\bar{n}$. Fig.2.c then shows that for $\bar{n}\sim$1/2 the fraction of excitons with no NN interaction is maximised, reaching around 90$\%$ for $P=$ 8 nW (Fig.2.d \fdd{and extended data Fig.4}), as expected for a CB solid. Increasing $P$ breaks this distribution, and remarkably for $\bar{n}\sim1$ we observe a characteristic feature of MIs, namely that the fraction of excitons interacting with 4 NN is maximised suddenly. For $P=$ 17 nW, it reaches around 90$\%$ of the exciton population (Fig.2.b \fdd{and extended data Fig.4}), underlining that excitons are spontaneously ordered with one exciton per site.

%

\fd{\textbf{Thermal melting of insulating phases:}} Overall, Fig.2 provides evidence for the theoretically expected MI and CB phases at $\bar{n}$=1 and $\bar{n}$=1/2 respectively. To further support this conclusion we measured the temperature dependence of the excitons compressibility. Hence, we study the thermal melting of the two insulating phases  and confront our observations with the theoretical phase diagram, measuring cuts along the horizontal axis in Fig.1.c (dotted lines).

For $\bar{n}=1$, Fig.3.a shows that $\kappa$ increases slowly towards the level given by Poissonian fluctuations (gray points). The latter is  reached for $T\gtrsim$ \SI{750}{\milli\kelvin}, in agreement with the theoretical critical temperature for the melting of the MI (vertical dashed line). In fact, the solid lines in Fig.3.a highlight that our observations accurately follow mean-field predictions, for both $\kappa$ and the level of Poissonian fluctations (violet and gray respectively). Hence, we confirm the calculated magnitudes of on-site interactions $U$, since these govern the melting of MIs \cite{Trivedi_2011,Gerbier_2007,deMArco_2005}.  This melting is possibly scrutinised by modelling the PL profile. Indeed, Fig.3.b-c signal that the fraction of excitons with no NN (orange) increases by around 20$\%$ between 330 and 750 mK. At the same time, the population of excitons occupying the 2$^{nd}$ WS or the continuum is also enhanced. These combined variations manifest that a significant fraction of empty sites is thermally activated. We attribute their emergence to the very strong on-site interaction strengths, so that thermal excitations expel excitons from the lattice (Fig.1.b). Finally, in the normal phase ($T\gtrsim$750 mK) we note that the various occupation fractions vary weakly.

On the other hand,  Fig.3.d reveals that for $\bar{n}=1/2$ $\kappa$ increases steeply while $T$ is enhanced. Below the critical temperature calculated for the melting of the CB phase, around \SI{400}{\milli\kelvin} (vertical dashed line), $\kappa$ is sub-poissonian and then follows the (classical) variation given by Poissonian fluctuations (gray points). Again, the solid lines in Fig.3.d signal that mean-field calculations quantitatively follow our experimental observations for both $\kappa$ and the Poissonian noise level (orange
and gray respectively). These predictions only rely on the Bose-Hubbard parameters calculated for the profile of our lattice potential, and by setting   $V_{1,1}=35$ $\mu$eV. This value lies in good agreement with the one deduced in Fig.2. Thereby we confirm that  $V_{1,1}$ is around 30 $\mu$eV, which is reasonable compared to the magnitude calculated from the theoretical profile of the lattice potential (Methods).

As for the MI phase, we finally studied the PL spectrum to extract the thermal variation of the exciton distribution for $\bar{n}=1/2$. At \SI{330}{\milli\kelvin}, Fig.3.e shows that the PL is only due to the recombination of excitons in the 1$^{st}$ WS with no NN interactions. Increasing $T$ we observe that the contribution at $\Delta$ above the lowest WS grows rapidly (middle panel of Fig.3.f). This higher energy PL, dominant for $T\gtrsim$ \SI{750}{\milli\kelvin},  reflects thermal excitations of the CB, so that excitons tunnel between lattice sites and then interact with excitons in neighbouring sites. Furthermore, the bottom panel of Fig.3.f verifies that both the 2$^{nd}$ WS and the continuum are weakly occupied, so that these states play a negligible role.

\fd{\textbf{Conclusions:}} \fd{Our studies evidence that dipolar excitons can be used to controllably implement extended Bose-Hubbard Hamiltonians. On the one hand this offers a new platform to map Ising models with bosonic arrays \cite{Broaweys_2021}, potentially across hundreds of lattice sites. On the other hand,}  for the strength of NN interactions extracted in our studies ($V_{1,1}\sim30$ $\mu$eV), we expect a lattice supersolid phase for exciton temperatures around \SI{10}{\milli\kelvin}, which is within experimental reach.  For our current device this phase would buildup in the lowest energy WS, but more exotic configurations seem accessible in the parameter space explorable with dipolar excitons. Particularly, for shallower lattices CB and lattice supersolids can theoretically form simultaneously in both $s$-like and $p$-like orbitals. Such multi-component symmetry breaking collective states would provide a novel realm for research of quantum matter.

\section*{Acknowledgments}
C.L. and F.D. would like to thank S. Suffit for support during sample fabrication, together with A. Reserbat-Plantey and B. Urbaszek for a critical reading of the manuscript. Work at CNRS was funded by IXTASE from the French Agency for Research (ANR-20-CE30-0032-01). The work at Princeton University (L.P. and K.B.) was funded by the Gordon and Betty Moore Foundation through the EPiQS initiative Grant GBMF4420, and by the National Science Foundation MRSEC Grant DMR 1420541. Research at ICFO acknowledges support from ERC AdG NOQIA; Agencia Estatal de Investigación (R$\&$D project CEX2019-000910-S, funded by MCIN/ AEI/10.13039/501100011033, Plan National FIDEUA PID2019-106901GB-I00, FPI, QUANTERA MAQS PCI2019-111828-2, QUANTERA DYNAMITE PCI2022-132919,  Proyectos de I+D+I “Retos Colaboración” QUSPIN RTC2019-007196-7); MCIN via European Union NextGenerationEU (PRTR);  Fundació Cellex; Fundació Mir-Puig; Generalitat de Catalunya through the European Social Fund FEDER and CERCA program (AGAUR Grant No. 2017 SGR 134, QuantumCAT \ U16-011424, co-funded by ERDF Operational Program of Catalonia 2014-2020); the computer resources and technical support at Barcelona Supercomputing Center MareNostrum (FI-2022-1-0042); EU Horizon 2020 FET-OPEN OPTOlogic (Grant No 899794); National Science Centre, Poland (Symfonia Grant No. 2016/20/W/ST4/00314); European Union’s Horizon 2020 research and innovation programme under the Marie-Skłodowska-Curie grant agreement No 101029393 (STREDCH) and No 847648  (“La Caixa” Junior Leaders fellowships ID100010434: LCF/BQ/PI19/11690013, LCF/BQ/PI20/11760031,  LCF/BQ/PR20/11770012, LCF/BQ/PR21/11840013). R.W.C. acknowledges support from the Polish National Science Centre (NCN) under Maestro Grant No. DEC2019/34/A/ST2/00081. \fd{All authors declare non-financial interest.}

\section*{\fd{Authors contributions}}
\fd{K.B. and L.P. realised the GaAs bilayer while C.L. and F.D. designed and fabricated the gate electrodes to realise the 250 nm period electrostatic lattice. C.L. and F.D. performed all experiments and data analysis. C.L., U.B., T.G, R.C., T.S., M.L., M.H. and F.D. contributed to the theoretical developments. All authors contributed to writing the manuscript. F.D. directed the project.}

\section*{Data availability}

Data supporting all the conclusions raised in this manuscript are available for download upon request.

\onecolumngrid

\newpage

\vspace{.5cm}

\centerline{\includegraphics[width=\linewidth]{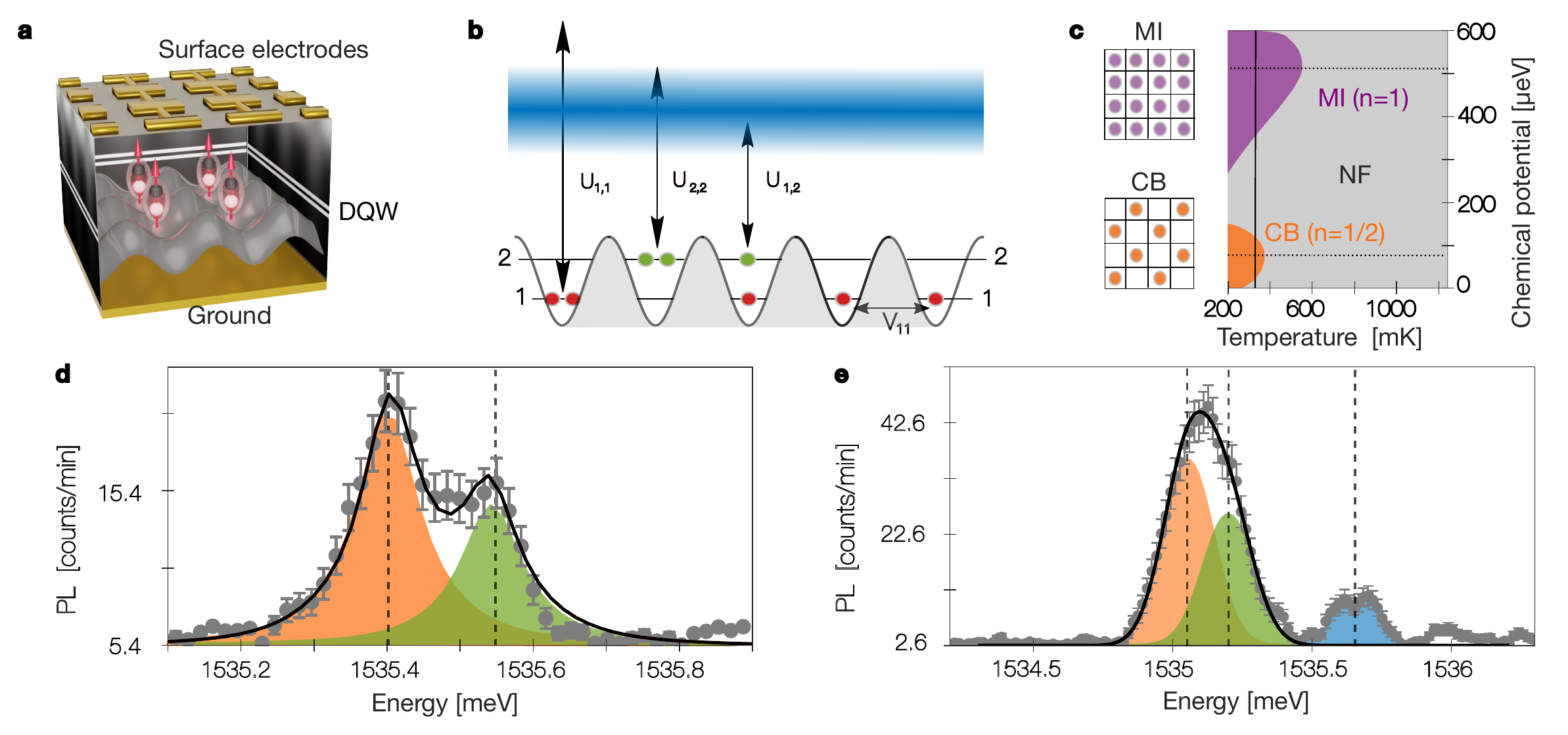}}
\textbf{Fig. 1}: \textbf{Strongly interacting dipolar excitons in a lattice.} \textbf{a} An array of polarized surface electrodes (gold) imprint a 250 nm period 2D lattice for dipolar excitons, made by electrons (gray) and holes (white) spatially separated in a double quantum well (DQW -- white layers). \textbf{b} In the lattice, dipolar excitons are confined in two WS (1 and 2). $U_{1,1}$ and $U_{2,2}$ denote on-site interaction strengths for excitons in WS 1 and 2 respectively, $U_{1,2}$ for excitons in distinct WS, while $V_{1,1}$ marks the strength of dipolar repulsions between NN sites for the 1$^{st}$ WS. \textbf{c} Phase diagram computed using a two-band mean-field model, highlighting that excitons can realise checkerboard (CB), Mott insulator (MI), and normal fluid (NF) phases. CB and MI configurations are illustrated on the left side while the vertical line highlights our lowest bath temperature. \textbf{d} PL spectrum for a mean density $\bar{n}$=0.2 exciton per site. The black line displays the modelled profile (R$^2$=98$\%$) by summing the emissions of excitons populating the WS 1 and 2, orange (67$\%$ fraction) and green (33$\%$) respectively. \textbf{e} PL spectrum for  $\bar{n}$=1.3 exciton per site. The solid black line provides the profile reproduced (with R$^2$=99$\%$) by setting 54$\%$ and 46 $\%$ occupations for the 1$^{st}$ and 2$^{nd}$ WS respectively (orange and green shaded regions). The blue area marks the contribution from excitons in the continuum states above the lattice. Measurements shown in \textbf{d}-\textbf{e} were performed at T=330 mK, error bars displaying Poissonian noise.

\newpage

\centerline{\includegraphics[width=.9\linewidth]{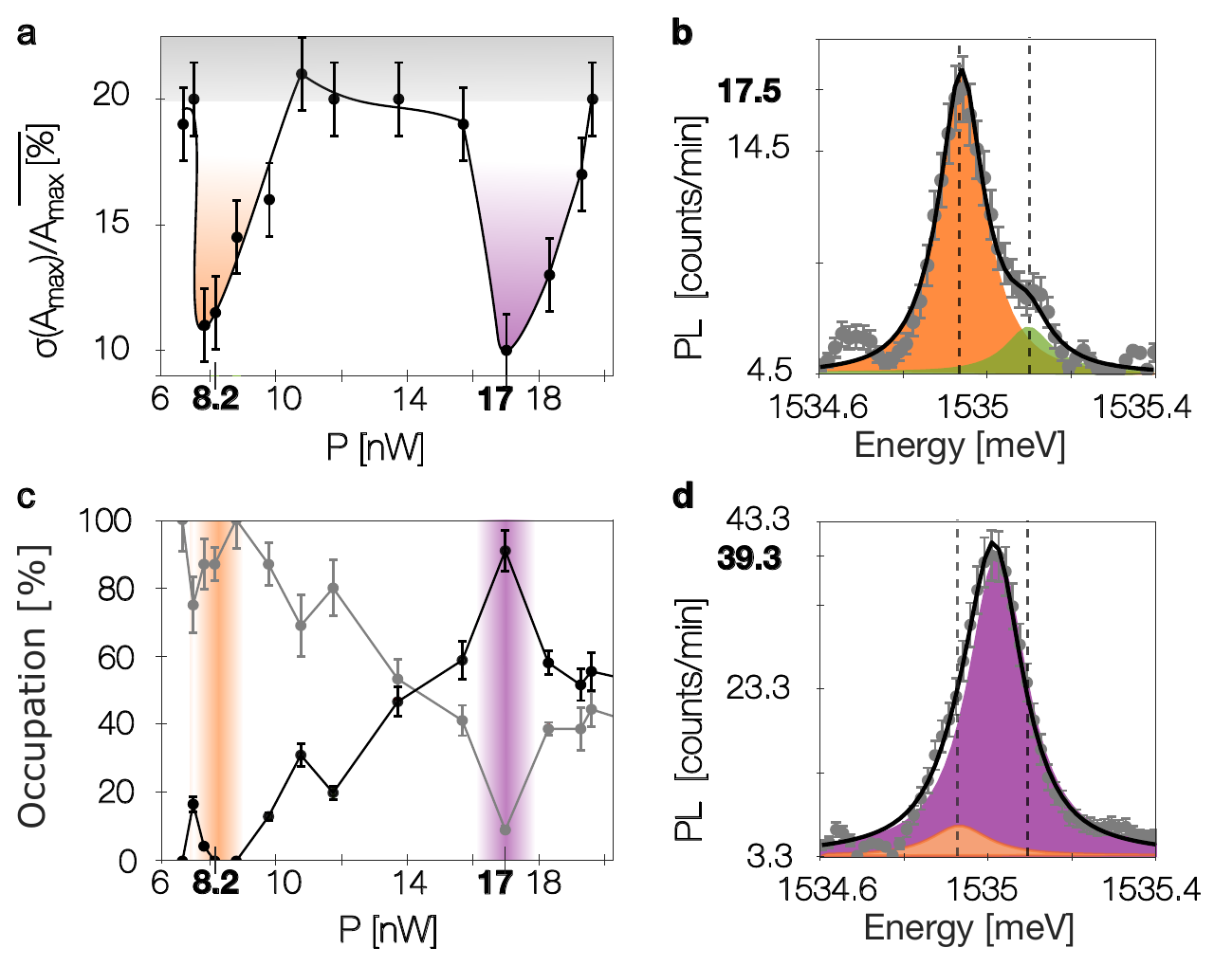}}
\textbf{Fig. 2}: \textbf{Incompressible phases at unity and half filling.}  \textbf{a} Fluctuations of the maximum of the PL spectrum $\sigma(A_\mathrm{max})/\overline{A_\mathrm{max}}$ as a function of the power $P$ of the laser excitation. \textbf{b} PL spectrum measured for $P$=8.2 nW, corresponding to $\bar{n}$=1/2. The black line shows the modelled profile (R$^2$=98$\%$) assigning a 90$\%$ fraction for excitons in the 1$^{st}$ WS (orange) and 10$\%$ in the 2$^{nd}$ WS (green).  \textbf{c} Occupation fractions of excitons occupying the 1$^{st}$ WS shifted in energy by $\Delta$ (due to four NN interactions -- black) and with no NN (gray), as a function of $P$. \textbf{d} PL spectrum measured for $P$=17 nW, corresponding to $\bar{n}$=1. The black line shows the modelled profile (R$^2$=99$\%$) assigning a 90$\%$ fraction  of excitons in the 1$^{st}$ WS with 4 NN (violet) and 10 $\%$ without NN (orange).  Measurements were all carried out at T=330 mK. Error bars provide our statistical confidence in \textbf{a}, the standard deviation in \textbf{c} and the poissonian noise in \textbf{b}-\textbf{d}.\vspace{.5cm}

\newpage

\centerline{\includegraphics[width=.9\linewidth]{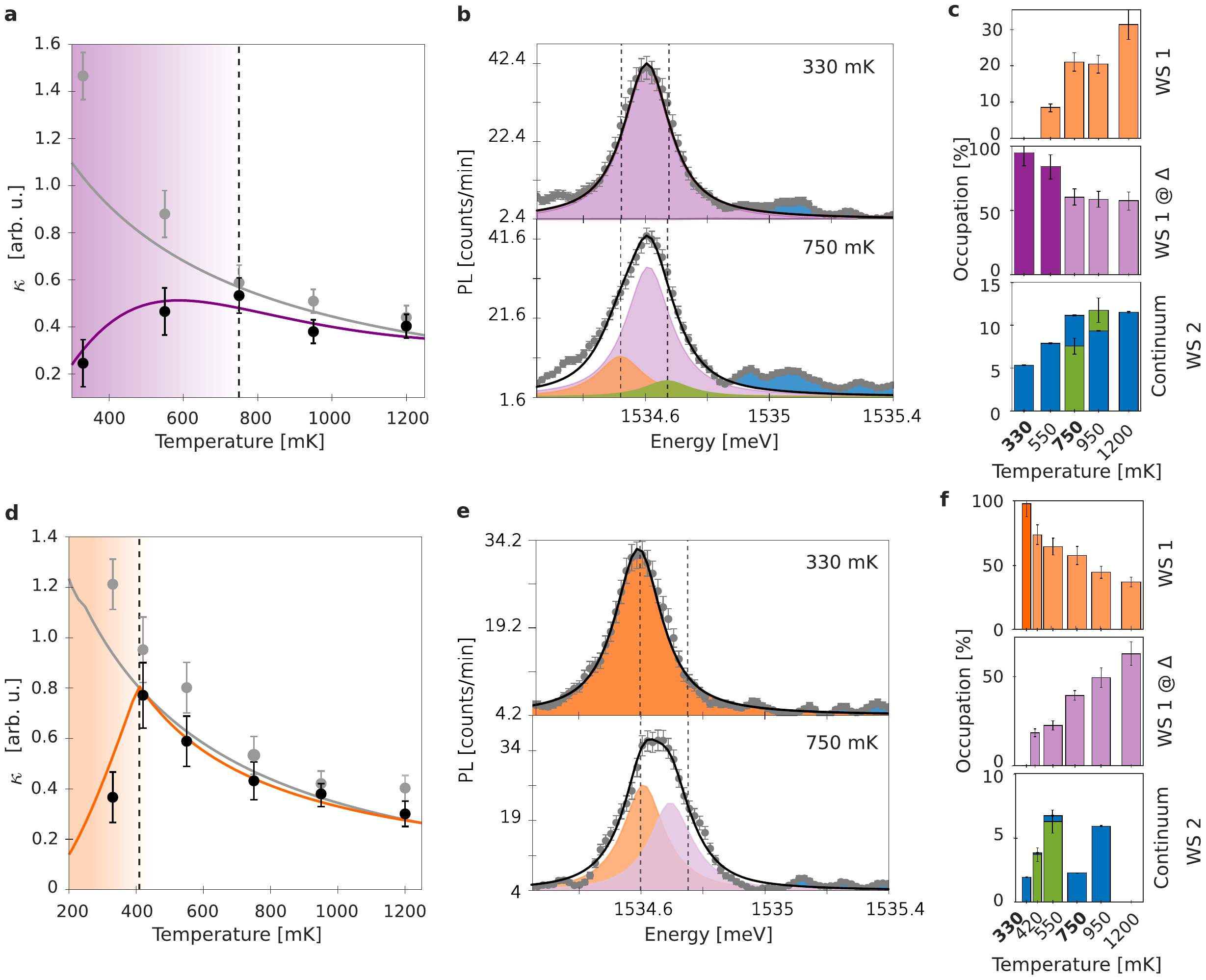}}
\textbf{Fig. 3}: \textbf{\fd{Melting of insulating phases.}} \textbf{a} Temperature variation of the compressibility $\kappa$ measured for $\bar{n}$=1 at the maximum of the PL spectrum (black points) and its theoretical variation deduced from mean-field calculations (violet line). Gray points and the gray line display the measured and theoretical level of poissonian fluctuations respectively.  \textbf{b} PL spectra at T=330 and 750 mK for $\bar{n}$=1. Black lines provide modelled profiles (R$^2$=98 and 97$\%$) with orange and violet areas marking the contributions from excitons in the 1$^{st}$ WS, without and with NN interactions. Excitons occupying the continuum (blue) and the 2$^{nd}$ WS (green) are also shown.  \textbf{c} Temperature dependence of the exciton fractions  occupying the 1$^{st}$ WS without (orange) or with (violet) NN interactions, together with the fraction of excitons occupying the continuum (blue) and the 2$^{nd}$ WS (green). \textbf{d}-\textbf{f} Same experimental results as in \textbf{a}-\textbf{c} but for $\bar{n}$=1/2. Error bars display our statistical precision in \textbf{a}-\textbf{d}, the poissonian noise level in \textbf{b}-\textbf{e} and the standard deviation in \textbf{c}-\textbf{f}.\vspace{.5cm}

\newpage

\twocolumngrid

\section*{Methods}

\subsection{Device and experimental procedure}

The \SI{250}{\nano\metre} period electrostatic lattice has been realized using an heterostructure described in Ref. \cite{Lagoin_2022}. It is based on two \SI{8}{\nano\metre} wide GaAs quantum wells, separated by a \SI{4}{\nano\metre} AlGaAs barrier. The quantum wells are positioned \SI{200}{\nano\metre} below the surface of the field-effect device where they are embedded, and \SI{150}{\nano\metre} above a conductive layer that serves as electrical ground. The lattice potential was engineered using the procedure detailed in the Supplementary Informations of Ref. \cite{Lagoin_2022}. Then, we designed and realized gate electrodes deposited at the surface of the heterostructure. These were polarised at around \SI{1}{\volt} in our experiments, yielding a periodically varying electric field perpendicular to the quantum wells plane. This field imprints the excitons lattice potential due to its interaction with the excitons permanent electric dipole. The latter has an amplitude around 12 $e.nm$, where $e$ denotes the electron charge, so that the lattice depth is about 250 $\mu$eV for 1 V applied across our 350 nm thick field-effect device.

In our studies electronic carriers are injected in the lattice potential using a laser excitation at resonance with the direct exciton absorption of each quantum well. The laser excitation is set with a rectangular profile at the surface of our device, with an area equal to around (10x5) $\mu m^2$.  We then study spectrally the PL reemitted by dipolar excitons in the (3x2.5) $\mu m^2$ central region (Figs.2-3), the horizontal extension being set by the width of our spectrometer's slit given our optical magnification. This area corresponds then to over 100 lattice sites. Moreover, PL spectra are acquired with a 1800 lines/mm grating. The PL is then sampled with 15 $\mu$eV precision so that the narrowest profile possibly measured, e.g. a laser line, is lorentzian-like with a 150 $\mu$eV full-width-at-half-maximum. In Fig.1-3, we assign this profile to the emission of the 1$^\mathrm{st}$ and 2$^\mathrm{nd}$ WS, as well as for the one of the 1$^\mathrm{st}$ WS shifted by $\Delta$. The energy position of each contribution is fixed throughout our analysis, so that we only adjust their amplitudes. Finally, note that the PL is acquired during a \SI{100}{\nano\second} long time interval, starting \SI{300}{\nano\second} after extinction of the loading laser pulse, while the excitons radiative lifetime exceeds \SI{700}{\nano\second}  \cite{Lagoin_2022}. Hence, we ensure that excitons are thermalised at the bath temperature and that the concentration of photo-injected excess carriers is minimised \cite{Alloing_2012,Beian_2017}.

To calibrate the average exciton density, as in Ref. \cite{Lagoin_2022}, we monitored the temporal dynamics of the PL energy following the laser excitation. For that we used a region without lattice potential so that dipolar excitons realize a homogeneous fluid. Comparing the PL energy detected \SI{300}{\nano\second} after the laser pulse, to the one at much longer delays, we deduce the magnitude of repulsive dipolar interactions that translates into the average exciton density. Thus, for $P\sim$ \SI{33}{\nano\watt} we observed that the energy shift is bound to \SI{150}{\micro\eV}, so that the exciton density is about \SI{6e9}{\centi\metre\squared}. In a \SI{250}{\nano\metre} period lattice potential this translates into $\bar{n}\sim$2 excitons per lattice site. Accordingly, we deduce that  $\bar{n}\sim$1 for $P=$ \SI{17}{\nano\watt}, since we verified that the average exciton density varies linearly with the power of the loading laser pulse.

\subsection{Physical parameters of the extended Bose-Hubbard Hamiltonian}

At low filling ($ \bar n \lessapprox 1$), we observe experimentally (e.g. Fig.3.e) that continuum states are barely occupied, such that our system is theoretically well described by a two-band extended BH model. In its most generic form, the corresponding Hamiltonian reads $H=\sum_i h_i + \sum_{\langle i,j\rangle} h_{ij}$, (see Ref. \cite{Dutta_2015}), with the on-site terms $h_i=\sum_{\alpha,\beta,\delta,\gamma} U_{\alpha\beta\delta\gamma} b_{i\alpha}^\dagger b_{i\beta}^\dagger b_{i\gamma} b_{i\delta}- \sum_\alpha \mu_\alpha n_{i\alpha}$, and the terms between NN sites $\langle i,j\rangle$ given by $h_{ij}= \sum_\alpha - t_\alpha (b_{i\alpha}^\dagger b_{j\alpha} + {\rm h.c.} ) + \sum_{\alpha,\beta,\delta,\gamma} V_{\alpha\beta\delta\gamma}  b_{i\alpha}^\dagger b_{j\beta}^\dagger b_{j\gamma} b_{i\delta}$, where Greek (Latin) indices represent band (site) degrees of freedom, $b_{i\alpha}$ ($b_{i\alpha}^\dagger$) are bosonic annihilation (creation) operators, and $n_{i\alpha}=b_{i\alpha}^\dagger b_{i\alpha}$. The dominant interaction terms are density-density interactions, $U_{1,1}\equiv U_{1111}$, $U_{2,2}\equiv U_{2222}$, and $U_{1,2} \equiv U_{1221}+U_{1212}+U_{2121}+U_{2112}$ for the on-site interactions, as well as $V_{1,1}\equiv V_{1111}$, $V_{2,2} \equiv V_{2222}$, and $V_{1,2}=V_{1221}+V_{2112}$ for NN interactions. Other interaction channels which mix bands have also been included within our calculations based on exact diagonalization (see Methods C), but are disregarded within the mean-field description (see Methods D).

To evaluate the strength of the different on-site and inter-site interactions, $U$ and $V$ respectively, we proceed as detailed in Refs. \cite{Lagoin_2022,Gotting_2022}. Relying on the spatial profiles of the Wannier wave-functions expected for our lattice potential with around 250 $\mu$eV depth, we estimate the magnitudes of $U$, $V$ and $t$. For the former, dominant terms are the density-density on-site interactions, namely $U_{1,1}\sim$ \SI{1}{\milli\eV} and $U_{2,2}\sim$ \SI{500}{\micro\eV} for the WS 1 and 2, which both exceed the excitons confinement depth in the lattice and can then not be measured. Also, we find that $U_{1,2}\sim$ \SI{200}{\micro\eV}. Density-density interactions are also the dominant inter-site interactions, and for the first WS our calculations yield $V_{1111} \equiv V_{1,1} \sim$ 15 $\mu$eV. This magnitude is two times smaller than the one deduced from our experiments. Nevertheless, in Fig.2-3, within the framework of the extended Bose-Hubbard model, we attribute the energy difference between MI and CB phases to dipolar interactions between excitons confined in NN sites only. Thereby we neglect longer-range contributions. These need to be included to possibly compare the value calculated from the lattice profile to the one extracted from our measurements. The value calculated for $V_{1,1}$ then has to be multiplied by a Madelung constant, of around 6/4, and is thus effectively increased from 15 to 22 $\mu$eV, in reasonable agreement with the measurements shown in Fig.2-3. Moreover, let us note that the lattice potential possibly confines excitons more weakly than expected. In this case, the overlap between Wannier wave-functions localized in neighbouring lattice sites would increase rapidly, resulting in larger values of $V_{1,1}$. Finally, for the tunnelling strength between nearest neighbouring lattice sites, our model calculations yield $t_1\sim$ \SI{1}{\micro\eV} and  $t_2\sim$ \SI{7}{\micro\eV} for the first and second WS respectively.

\subsection{Exact diagonalisation calculations}

For the theoretical description we consider the extended two-band Bose-Hubbard Hamiltonian, relying on the theoretically expected parameters for $t_\alpha$, $U_{\alpha\beta\gamma\delta}$, and $V_{\alpha\beta\gamma\delta}$, but with increased amplitude for density-density inter-site interactions to match the experimentally measured value of $V_{1,1}$ in the first WS. Specifically, we use $V_{1,1}=35 \mu$eV, $V_{2,2}= 250 \mu$eV and $V_{1,2}= 40 \mu$eV.

We obtain the full eigen-spectrum of the Hamiltonian by applying exact diagonalization (ED) on a supercell, spanned by ${\bf L}_1=(2,2)a$ and ${\bf L}_2=(2,-2)a$, $a$ denoting the lattice period, which contains 8 sites of the square lattice. In the ED calculation, we fix the particle number to 4 excitons (half filling). From the eigenspectrum, we calculate thermal expectation values of observables of interest. A hallmark of CB order is a peak of the structure factor at ${\bf k}_{\rm CB} \equiv (\frac{\pi}{a},\frac{\pi}{a})$. Indeed, the structure factor in the first WS, defined as $S_1({\bf k}) \sim \sum_{ij} (\langle n_{i1}n_{j,1}\rangle -\langle n_{i1}\rangle \langle n_{j1}\rangle) e^{-i {\bf k}\cdot {\bf R}_{ij}}$, with ${\bf R}_{ij}$ the lattice vector connecting sites $i$ and $j$, is found to exhibit a pronounced peak at ${\bf k}_{\rm CB}$ (\fd{extended data Fig.1.a}), with a value $|S_1({\bf k}_{\rm CB})|$ which remains more than twice as large as any other value of $|S_1({\bf k})|$ up to temperatures as large as $T\sim 400$ mK (\fd{extended data Fig.1.b}). 

\subsection{Mean-field calculations}

Larger system sizes are studied in the mean-field approximation which reduces the Hamiltonian to a sum of single-site terms, $H^{\rm MF} =\sum_i (h_i + h_{i}^{\rm NN})$, with $h_{i}^{\rm NN}= \sum_{j} [-\sum_{\alpha} t_\alpha (b_{i\alpha}^\dagger \langle b_{j\alpha} \rangle - \langle b_{i\alpha}^\dagger \rangle  \langle b_{j\alpha} \rangle  +{\rm h.c.}) + \sum_{\alpha,\beta} V_{\alpha\beta\beta\alpha} (n_{i\alpha} \langle n_{j\beta} \rangle - \langle n_{i\alpha} \rangle \langle n_{j\beta} \rangle].$ The sum in $j$ includes all nearest neighbouring sites of $i$, and for simplicity we restrict ourselves to density-density interactions. On every site, the 4 mean-field values $\langle b_{i1}\rangle$, $\langle b_{i2}\rangle$, $\langle n_{i1}\rangle$, and $\langle n_{i2}\rangle$ have to be chosen such that they self-consistently match the corresponding thermal expectation values obtained from the solution of the mean-field Hamiltonian. In our numerical calculation, the self-consistency loop sweeps through a 4x4 lattice. From the solution of the self-consistent mean-field Hamiltonian, we calculate the phase diagram, shown in Fig.1.c, and the compressibility $\kappa$, shown in Fig.3. Fixing the chemical potential such that the filling corresponds to $\bar n=1$ or $\bar n= 1/2$, the temperature-dependence of $\kappa$ shown in Figs. 3.a and 3.d match quantitatively the experimental data. Moreover, \fd{extended data Fig.1.c} plots the population imbalance between sublattices whose non-zero values theoretically characterize the CB phase.

\onecolumngrid

\pagebreak

\newpage

\pagebreak

\centerline{\includegraphics[width=\linewidth]{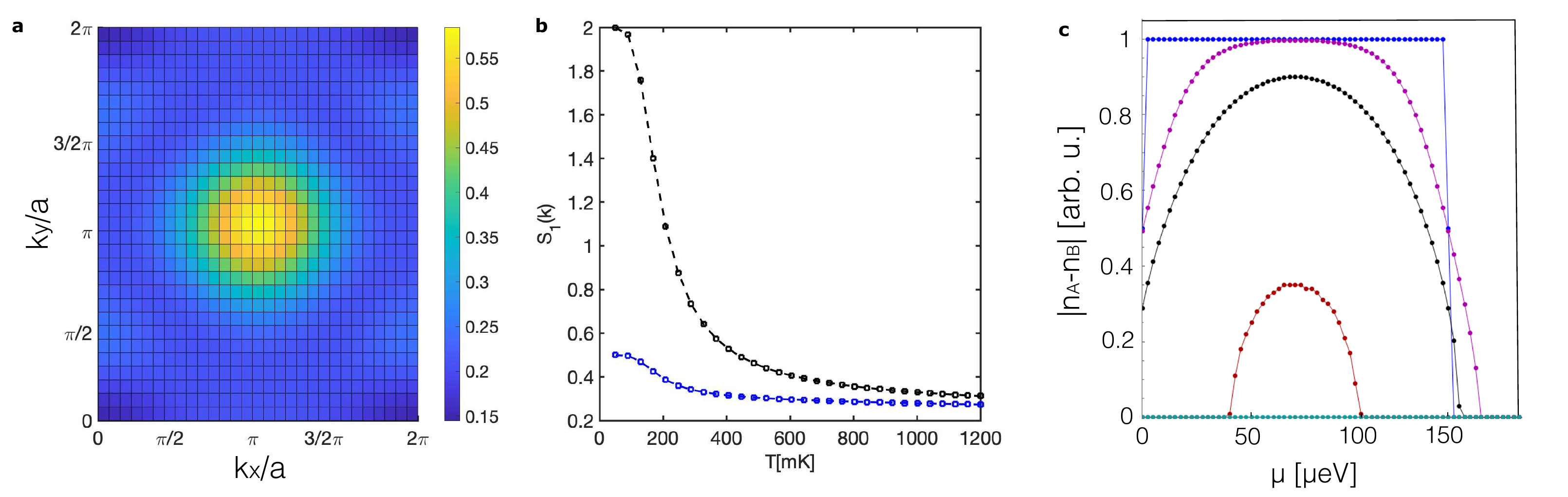}}
\fd{\textbf{Extended Data Fig.1}:} \textbf{Theoretical hallmarks of CB order.}
\textbf{a} Lowest band structure factor $S_1({\bf k})$ at T=100 mK obtained by exact diagonalisation of a 8 site square lattice (Betts cluster) with periodic boundary conditions. It exhibits a dominant peak at quasi-momentum ${\bf k}=(\pi/a,\pi/a)$, which is a characteristic  signature of CB order.  A second strongly suppressed quasi-peak lies at ${\bf k}=(0,0)$ (due to finite size effects), corresponding to a homogenous liquid without any density order. \textbf{b} $|S_1(\pi/a,\pi/a)|$ (black) and $|S_1(0,0)|$ (blue) are plotted vs. temperature $T$. Up to $T\lesssim T_c=$ 420 mK, the structure factor signalling CB order remains at least twice as large as the structure factor for a homogeneous liquid. \textbf{c} CB order parameter deduced from mean-field calculations as a function of the chemical potential $\mu$ and temperature ($T=$ 4, 125, 247, 389, 450 mK in blue, violet, black, red and green respectively). The order parameter is given by the population difference $|n_A-n_B|$ between two sub-lattices, $A$ and $B$, of the square lattice. Below around 410 mK $|n_A-n_B|$ is significant manifesting CB order.\vspace{.5cm}

\newpage

\centerline{\includegraphics[width=.8\linewidth]{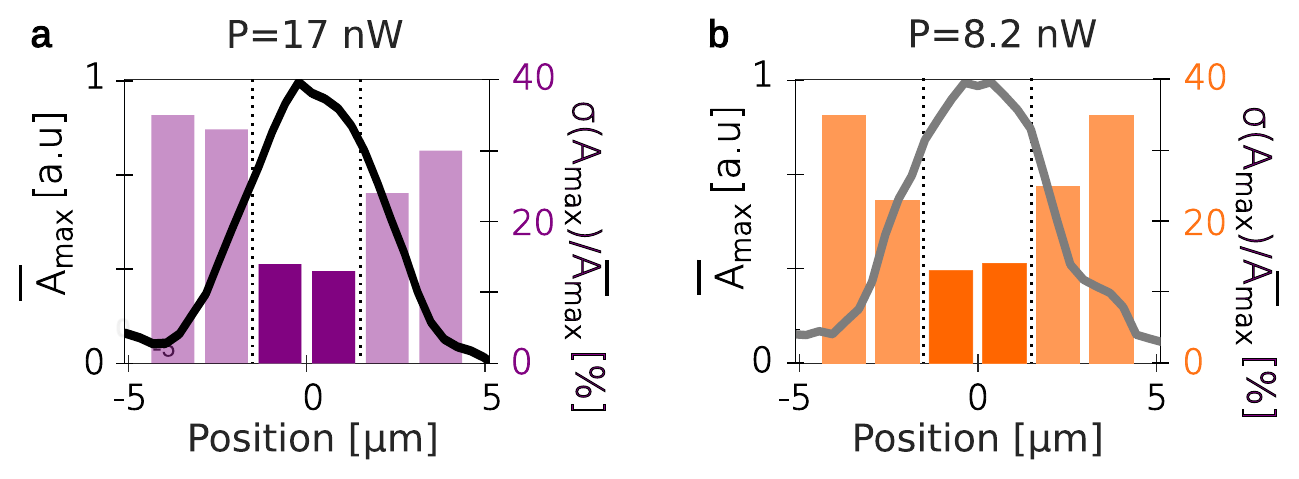}}
\fd{\textbf{Extended Data Fig.2}}: \textbf{Spatially resolved PL intensity and intensity fluctuations.} \textbf{a} Spatial variations of the PL intensity $\overline{A_\mathrm{max}}$ (black line) and $\sigma(A_\mathrm{max})/\overline{A_\mathrm{max}}$ (violet bars) measured at $T=330$ mK and $P=17$ nW, i.e. for the MI phase. Both $\overline{A_\mathrm{max}}$ and $\sigma(A_\mathrm{max})/\overline{A_\mathrm{max}}$ vary weakly in the 3 $\mu$m central region of the laser excited region, evidencing that the MI phase is homogeneous across over 100 lattice sites. Outside this region we note that $\sigma(A_\mathrm{max})/\overline{A_\mathrm{max}}$ increases steeply while $\overline{A_\mathrm{max}}$ drops, which signals that excitons realise a normal fluid.  \textbf{b} Same measurements obtained for $P=8.2$ nW, i.e. for the CB phase. Results are extracted from the experiments reported in Fig.2.\vspace{.5cm}

\newpage

\centerline{\includegraphics[width=.6\linewidth]{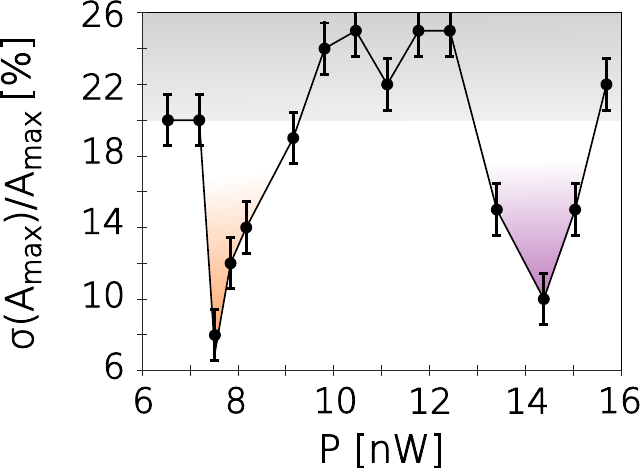}}
\fd{\textbf{Extended Data Fig.3}}: \textbf{Exciton compressibility vs. average lattice filling.} \textbf{a} Fluctuations of the maximum of the PL intensity ($\sigma(A_\mathrm{max})/\overline{A_\mathrm{max}}$) as a function of the power $P$ of the loading laser, in a different region of our two-dimensional square lattice. As for Fig.2, experimental results are obtained by statistically analysing a series of 10 measurements for every value of $P$. The laser excitation profile was set close to the one for the experiments shown in Fig.2. Remarkably we recover that two insulating phases emerge for $P=$ 7 and 14.4 nW, in good agreement with the findings discussed in the main text. Experiments were realised at $T=$ 330 mK, error bars display statistical confidence while the level of Poissonian fluctuations is given by the gray shaded region.\vspace{.5cm}

\newpage

\centerline{\includegraphics[width=.7\linewidth]{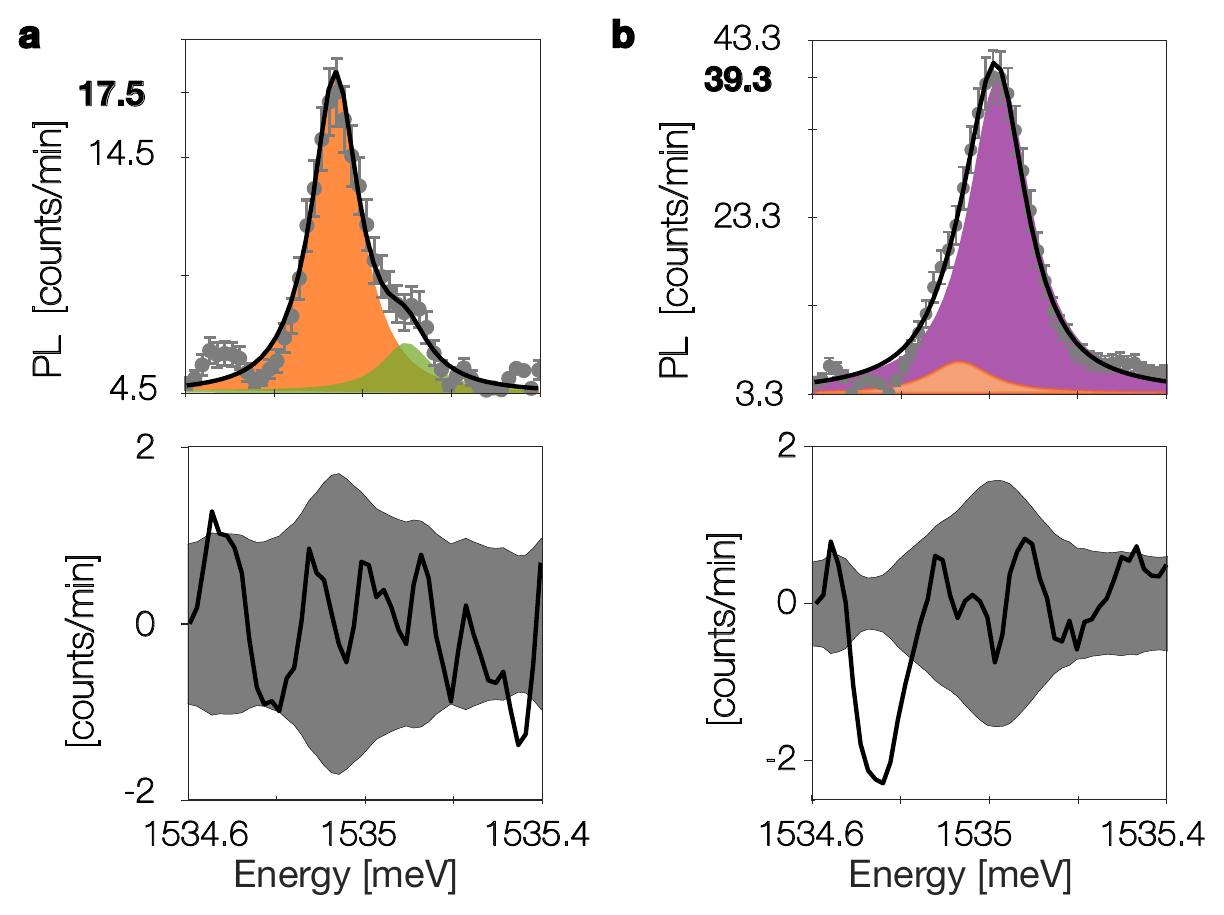}}
\fd{\textbf{Extended Data Fig.4}}: \textbf{Residuals at $\bar{n}=$ 1/2 and 1.} \textbf{a} PL spectrum measured at $\bar{n}$=1/2 (top) together with the modelled profile (black line). The bottom panel displays the residuals between modelled and measured profiles (black line), compared to the amplitude of poissonian fluctuations (gray area). \textbf{b} Same measurements for $\bar{n}$=1. Experimental results are taken from the data reported in Fig.2.b-d. 

\end{document}